\documentclass[showpacs,preprintnumbers,amsmath,amssymb,11pt]{article}
\usepackage[utf8]{inputenc}
\usepackage[italian,english]{babel}
\usepackage{amsmath}
\usepackage{graphicx}
\usepackage[colorinlistoftodos]{todonotes}
\usepackage{caption}
\usepackage{subcaption}
\usepackage{tabularx}
\usepackage{bbm}
\usepackage{listings}
\usepackage{amsthm}
\usepackage{algorithm}
\usepackage{algpseudocode}
\usepackage{mathtools}
\usepackage{amsmath}
\usepackage[toc,page]{appendix}
\usepackage{booktabs}
\usepackage{xcolor}
\usepackage{amsfonts}
\usepackage{epstopdf}
\usepackage{adjustbox}
\usepackage{csquotes}

\usepackage[numbers]{natbib}

\DeclareMathOperator*{\argmin}{arg\,min}

\makeatletter
\def\BState{\State\hskip-\ALG@thistlm}
\makeatother

\definecolor{darkgreen}{rgb}{0,0.5,0}
\definecolor{darkred}{rgb}{0.5,0,0}

\usepackage{amsmath,amsfonts,amssymb,amsthm,epsfig,epstopdf,titling,url,array}

\theoremstyle{plain}
\newtheorem{thm}{Theorem}[section]
\newtheorem{lem}[thm]{Lemma}
\newtheorem{prop}[thm]{Proposition}
\newtheorem{cor}[thm]{Corollary}

\theoremstyle{plain}
\newtheorem{defn}{Definition}[section]

\theoremstyle{remark}

\newtheorem*{note}{Note}

\newcommand{\E}{\mathbb{E}}
\newcommand{\pdf}{\textit{pdf}}
\newcommand{\chf}{\textit{chf}}

\newcommand{\mycite}[1]{~\cite{#1}}

\newcommand{\rv}{\textit{rv}}
\newcommand{\id}{\textit{id}}

\newcommand{\sd}{\textit{sd}}

\newcommand{\eqd}{\stackrel{d}{=}}

\newcommand{\arem}{$a$-remainder}

\newcommand{\BM}{\emph{BM}}

\newcommand{\Levy}{L\'{e}vy}

\title{Correlating Lévy processes with Self-Decomposability: Applications to Energy Markets
\thanks{The views, opinions, positions or strategies expressed in this work are those of the authors and do not represent the views, opinions and strategies of, and should not be attributed to E.ON SE.}
}

\author{
	Matteo Gardini\thanks{Department of Mathematics, University of Genoa, Via Dodecaneso 16146, Genoa, Italy, email gardini@dima.unige.it}
	\and
	Piergiacomo Sabino\thanks{Quantitative Modelling E.ON SE
		Br\"usseler Platz 1, 45131 Essen, Germany, email piergiacomo.sabino@eon.com}
	\and
	 Emanuela Sasso\thanks{Department of Mathematics, University of Genoa, Via Dodecaneso 16146, Genoa, Italy, email sasso@dima.unige.it}}

\date{\today}
\begin{document}
\maketitle

\begin{abstract}
Based on the concept of self-decomposability, we  extend some recent multivariate \Levy\ models built using multivariate subordination with the aim of capturing  situations in which a sudden event in one market is propagated onto related markets after a certain stochastic time delay.

Consequently, we study the properties of such processes, derive closed form expressions for the characteristic function and detail how a Monte Carlo scheme can be easily implemented. 

We illustrate the applicability of our approach in the context of gas and power Energy markets focusing on the calibration and on the pricing of spread options written on different underlying assets using simulations techniques.

\vspace{0.2cm}
\end{abstract}

\section{Introduction}
\label{sec:Intro}
During the last decades a lot of efforts have been done to go beyond the \citet{BLS1973} framework in Financial Modelling. The Black-Scholes (BS) formula is widely used by practitioners but its limits are well-known. Over the years a lot of researchers - \citet{Merton76}, \citet{MadanSeneta90} and \citet{Heston93} among others - have proposed more sophisticated models to overcome its limitations. Nevertheless, their focus is mainly on the single asset modelling framework.
\par If a multi-asset market has to be considered one has to take care about modeling the dependence structure and this can be a tricky task. One mainly comes up against three issues: 
\begin{itemize}
	\item How to extend a univariate model to a multivariate setting preserving mathematical tractability?
	\item How to calibrate this model?
	\item Which techniques can be used for derivatives pricing?
\end{itemize}
Beyond the Gaussian world, some choices have been proposed to model dependence in the context of \Levy\ processes. Among others, \citet{CT2003}, \citet{CLV2013}, \citet{Panov2019} and \citet{Panov2015} have discussed the use of \Levy\ copulas or \Levy\ series representation. Unfortunately, these approaches, especially \Levy\ copulas, are difficult to handle mathematically and are often hard to calibrate. 

In this study we address the three issues above in the context of multi-dimensional processes, that are at least 
marginally \Levy, using multivariate subordination. To this end, several approaches are available in the literature, for instance 
\citet{BNPS2001} or \citet{LS2006} use a common subordinator. In particular, in a 
series of papers \citet{Semeraro2008}, \citet{SL2010}, \citet{BB2013}, \citet{Buchman2017} and \citet{Buchman2019} have proposed models 
based on subordination to introduce dependence between \Levy\ process. The common idea of these papers is to define multivariate 
processes that are the sum of an independent process and a common one. For example \citet{BB2013} define a multivariate process in 
the following way:
\begin{equation*}
\boldsymbol{Y}\left(t\right)= \left(Y_{1}\left(t\right),\dots, Y_{n}\left(t\right)\right)^{T} = \left(X_{1}\left(t\right) + a_{1}Z\left(t\right),\dots, X_{n}\left(t\right) + a_{n}Z\left(t\right)\right)^{T}
\end{equation*}
where $Z\left(t\right)$, $X_{j}\left(t\right)$, $j=1,\dots,n$ are independent \Levy\ processes. In a financial market, one can  see the 
 common process $Z\left(t\right)$ as a systemic risk, whereas the independent processes $X_{j}\left(t\right)$ can be considered as an idiosyncratic component. The model has a simple economical interpretation and it is mathematically tractable. 
 
 The assumption that the systemic risk is a driven by a common process $Z\left(t\right)$ simplify the modeling approach but on the other hand, specially in illiquid markets, can be too narrow.
  Indeed, cases in which we observe delays in market reactions are not so rare. Sometimes a general event has an effect on a market but others related markets could not immediately react. Anyway, it can happen that, as the time goes on, other related markets can be influenced from such an event. As matter of fact we observe a sort of \enquote{delay in the propagation of the information} across markets and its clear that such a situation is not taken in account by the existing models. 
 
 The aim of this paper is to use the notion of \sd, following the approach proposed by \citet{cs20}, to extend previous existing models presented by \citet{Semeraro2008}, \citet{SL2010} and \citet{BB2013} so that the \enquote{delay in innovations propagation} effect is considered. 
     This last feature can be captured by simply adding one parameter to the approaches  mentioned above  without implying a remarkable model complication.
 From a mathematical point of view it is also worthwhile observing
that our model goes beyond the mathematical generalization of the original ones provided by \citet{Buchman2017} and
\citet{Buchman2019}: the authors analyze the case where the subordinator is \sd. As it will be clear from the sequel, the
$a$-reminder part of the subordinator process is infinitely divisible but not \sd. 
\par Looking at calibration issue, general techniques, such as Non-Linear-Least-Square (NLLS) or Generalized Method of Moments (GMM), can be adapted to our case, leading to a two-step calibration method as the one presented by \citet{BB2013}.
\par About derivative pricing, since \chf's are know in closed form, methods based on Fourier transform, as the ones presented by \citet{hurd2009}, \citet{Pellegrino2016} and \citet{caldanafusai2016}, can be applied. Moreover standard numerical schemes for path simulations can be adapted to our model, leading to numerical pricing techniques based on Monte Carlo simulations. \\

The article is organized as follow: in Section  \ref{sec:Prelim} we give the basic notions that we need in the sequel and we point up an economic interpretation of proposed modeling framework. In Sections
\ref{sec:modExtension} we detail how to extend the  models of \citet{Semeraro2008},  \citet{SL2010} and \citet{BB2013} using \sd\
subordinators, whereas in Section \ref{sec:NumRes} we briefly outline avaiable calibration and pricing techniques, we calibrate
these models on Power and Gas Forward markets and we price spread options. Section \ref{sec:Conclusions} concludes the paper. All proofs are given in Appendix \ref{sec:AppendixA}.

\section{Preliminaries}
\label{sec:Prelim}
In this section we introduce the fundamental concepts we need in the sequel: \sd\ laws and Brownian subordination. We look at \sd\ as a natural way to generate correlated \rv\ and we use this notion to build dependent stochastic processes in continuous time. We define increasing dependent stochastic processes and we use the subordination technique to build dependent subordinated Brownian Motions (\BM). We refer to \citet{CT2003}, \citet{Sato} and \citet{Cufaro08} for the details.

We recall that a law with probability density (\pdf) $f(x)$ and
characteristic function (\chf) $\varphi(u)$ is said to be
\emph{self-decomposable} (\sd) (see Sato\mycite{Sato} or Cufaro
Petroni~\cite{Cufaro08}) when for every $0<a<1$ we can find another
law with \pdf\ $g_a(x)$ and \chf\ $\chi_a(u)$ such that
                \begin{equation}\label{aremchf}
                    \varphi(u)=\varphi(au)\chi_a(u).
                \end{equation}
We will accordingly say that a random variable (\rv) $X$ with \pdf\
$f(x)$ and \chf\ $\varphi(u)$ is \sd\ when its law is \sd: looking
at the definition, this means that for every $0<a<1$ we can always
find two \emph{independent} \rv's, $Y$ (with the same law of $X$)
and $Z_a$ (here called \emph{\arem}), with \pdf\ $g_a(x)$
and \chf\ $\chi_a(u)$ such that
\begin{equation}\label{sdec-rv}
    X\eqd aY+Z_a.
\end{equation}

It is easy to see that $a$ plays the role of correlation coefficient between $X$ and $Y$: from here follows the idea is to build stochastic \Levy\ processes starting from \rv\ with \sd\ laws. To this end, it is well-known that if a law is \sd\ then is infinitely divisible (\id) and for a given $a\in\left(0,1\right)$ the law of $Z_{a}$ is uniquely determined and \id\ (see \citet[Proposition~15.5]{Sato}).
Since the laws of $X,Y$ and $Z_{a}$ have \id\ laws then we can construct the associated \Levy\ process (\citet[Proposition~3.1]{CT2003}).\\
Other important concepts are the notions of subordinators, that are almost surely non-decreasing \Levy\ processes, and Brownian subordination
(see \citet{CT2003}). One can use a non-decreasing \Levy\ process, called subordinator, $G\left(t\right)$ to \emph{time-change} a \Levy\ process obtaining a new one (\citet[Theorem~4.2]{CT2003}). If the time-change is done on a \BM\ this operation is then called Brownian subordination.

\begin{defn}
Consider a probability space $\left(\Omega,\mathcal{F},\mathbb{P}\right)$,  $\mu \in \mathbb{R}$ and $\sigma \in \mathbb{R}^{+}$.  Let $W\left(t\right)$ be a \BM\ and let $G\left(t\right)$ be a subordinator. A subordinated \BM\ $X\left(t\right)$ with drift is defined as:
\begin{equation}
X\left(t\right) = \mu G\left(t\right) + \sigma W\left(G\left(t\right)\right)
\label{eqn:subProcess}
\end{equation}
\end{defn}

\noindent If $H$ is $\mathbb{P}$-\emph{a.s}. non-negative random variables with \sd\ law we can build \sd\ subordinators as follows

\begin{defn}[Self-decomposable subordinators]
Let $\tilde{H}_{1}$ and $\tilde{H}_{2}$ be $\mathbb{P}$-\emph{a.s}. non-negative \rv\ with \sd\ laws and define $H_{i}\left(t\right)$ and $Z_{a}\left(t\right)$ as \Levy\ processes such that  $\left(H_{i}\left(1\right)\right) \eqd \left(\tilde{H}_{i}\right), i=1,2$ and $Z_{a}\left(1\right) \eqd \tilde{Z}_{a}$.
\textit{\sd\ subordinators} are defined as:

\begin{equation}
H_{2}\left(t\right) = a H_{1}\left(t\right) + Z_{a}\left(t\right)
\label{eqn:subordinatorsSD}
\end{equation}
\end{defn}

\noindent Note that the process $H_{2}\left(t\right)$ defined in \eqref{eqn:subordinatorsSD} is a Lévy process because it is a linear combination of two Lévy processes (\citet[Theorem~4.1]{CT2003}). 

\begin{figure}[ht]
    \centering
    \includegraphics[scale=0.3]{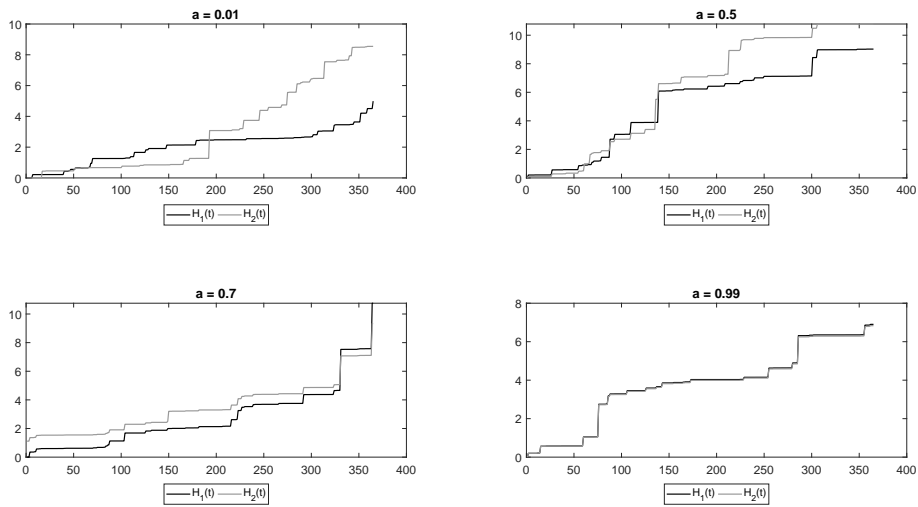}
    \caption{Correlated subordinators $H_{1}\left(t\right)$ and $H_{2}\left(t\right)$ with different values of $a$.}
    \label{fig:sdsubordinators}
\end{figure}

The construction proposed by Equation \eqref{eqn:subordinatorsSD} has a clear financial interpretation. Stochastic times processes $H_{1}\left(t\right),H_{2}\left(t\right)$ \enquote{run together} with a stochastic delay, given by the parameter $a$ and by the term $Z_{a}\left(t\right)$, one with respect to the other. In Figure \ref{fig:sdsubordinators} different paths of the process $\boldsymbol{H}\left(t\right) = \left(H_{1}\left(t\right),H_{2}\left(t\right)\right)$ are shown, varying the parameter $a$: for fixed $t$ the difference between $H_{1}\left(t\right)$ and $H_{2}\left(t\right)$ can be viewed as stochastic delay. Roughly speaking one can observe if $a \to 1$ then $H_{1}\left(t\right)$ and $H_{2}\left(t\right)$ are essentially indistinguishable.

This construction provides us a powerful tool to model those markets in which, whenever an event occurs in one of them, the effect on the other ones is not immediate but it occurs with a certain time delay.
Observe that the parameter $a$ is the only parameter we have to add to include this feature in our model and this do not leads to a significant model complication.

The former construction can be extended to the case $n >2$.\\
Define $\boldsymbol{H}\left(t\right) = \left(H_{1}\left(t\right),\dots,H_{n}\left(t\right)\right)$, $n\in \mathbb{N}$ by setting:
\begin{align*}
& H_{1}\left(t\right) \\
& H_{2}\left(t\right) = a_{1}H_{1}\left(t\right) + Z_{a_{1}}\left(t\right)\\
& \cdots \\
& H_{n}\left(t\right) = a_{n-1}H_{n-1}\left(t\right) + Z_{a_{n-1}}\left(t\right)\\
\end{align*}
where $\left(a_{j}\right)_{j=1}^{n-1} \in \left(0,1\right)$.

\par In next sections we extend some recently proposed multivariate \Levy\ models using the \sd\ subordinator $\boldsymbol{H}\left(t\right)$. Hereafter, for the sake of notational simplicity, we consider only the case with $n=2$.

\section{Model extensions with Self-Decomposability}
\label{sec:modExtension}
In this section we extend the models presented by \citet{Semeraro2008}, \citet{SL2010} and \citet{BB2013} using \sd\ subordinators introduced in Section \ref{sec:Prelim}.
\subsection{Extension of Semeraro's Model}
\label{sec:SSD}
The first model we extend using \sd\ subordinators was proposed by \citet{Semeraro2008}.

\begin{defn}[\sd-Semeraro Model]
Let $I_{j}\left(t\right) \; j=1,2$ be independent subordinators, and $H_{1}\left(t\right)$,  $H_{2}\left(t\right)$ be \sd\ subordinators defined in Equation \eqref{eqn:subordinatorsSD}, independent of $I_{j}\left(t\right)$. Define the subordinator $G_{j}\left(t\right)$

\begin{equation}
G_{j}\left(t\right) = I_{j}\left(t\right) + \alpha_{j}H_{j}\left(t\right), \quad j=1,2
\label{eqn:SSubordinators}
\end{equation}
with $\alpha_{j} \in \mathbb{R}^{+}$. Let $\mu_{j} \in \mathbb{R}$, $\sigma_{j} \in \mathbb{R}^{+}$, $W_{j}\left(t\right)$ be standard independent BM's and let $G_{j}\left(t\right)$ subordinators as is \eqref{eqn:SSubordinators}. Define the subordinated \BM\ with drift $Y_{j}\left(t\right)$ as:
\begin{equation}
Y_{j}\left(t\right) = \mu_{j}G_{j}\left(t\right) + \sigma_{j}W_{j}\left(G_{j}\left(t\right)\right), \quad j=1,2.
\label{eqn:subordinatedprocess}
\end{equation}
\end{defn}
Observe that the \enquote{delay in time effect} appears at the level of subordinators $G_{j}\left(t\right)$ and it is given by the couple $\left(H_{1}\left(t\right),H_{2}\left(t\right)\right)$.\\


The joint \chf\ of the process defined in \eqref{eqn:subordinatedprocess} has a nice closed expression.

\begin{prop}[Characteristic Function]
\label{prop:jointChfSemeraro}
The joint \chf\  $\phi_{\boldsymbol{Y}\left(t\right)}\left(\boldsymbol{u}\right)$ of the process $\boldsymbol{Y}\left(t\right) = \left(Y_{1}\left(t\right),Y_{2}\left(t\right)\right)$ at time $t$ defined in \eqref{eqn:subordinatedprocess} is given by:

\begin{equation}
\begin{split}
\phi_{\boldsymbol{Y}\left(t\right)}\left(\boldsymbol{u}\right) = &\phi_{I_{1}\left(t\right)}\left(u_{1}\mu_{1} + i \frac{\sigma_{1}^{2} u_{1}^{2}}{2}\right)
\phi_{I_{2}\left(t\right)}\left(u_{2}\mu_{2} + i \frac{\sigma_{2}^{2} u_{2}^{2}}{2}\right)
\phi_{Z_{a}\left(t\right)}\left(u_{2}\mu_{2} + i \frac{\sigma_{2}^{2} u_{2}^{2}}{2}\right) \\
& \phi_{H_{1}\left(t\right)}\left(\alpha_{1}\left(u_{1}\mu_{1} + i \frac{\sigma_{1}^{2} u_{1}^{2}}{2}\right) + a \alpha_{2}\left(u_{2}\mu_{2} + i \frac{\sigma_{2}^{2} u_{2}^{2}}{2}\right)\right)
\end{split}
\label{eqn:generalchf}
\end{equation}
\end{prop}

\begin{note}
Observe that the derived model is an extension of the one presented by \citet{Semeraro2008}. 
By taking the limit for $a\to 1$ in \eqref{eqn:generalchf} we have that:
\begin{equation*}
\begin{split}
\lim_{a\to 1} \phi_{\boldsymbol{Y}\left(t\right)}\left(\boldsymbol{u}\right) = &\phi_{I_{1}\left(t\right)}\left(u_{1}\mu_{1} + i \frac{\sigma_{1}^{2} u_{1}^{2}}{2}\right)
\phi_{I_{2}\left(t\right)}\left(u_{2}\mu_{2} + i \frac{\sigma_{2}^{2} u_{2}^{2}}{2}\right)
\\
& \phi_{H_{1}\left(t\right)}\left(\alpha_{1}\left(u_{1}\mu_{1} + i \frac{\sigma_{1}^{2} u_{1}^{2}}{2}\right) + \alpha_{2}\left(u_{2}\mu_{2} + i \frac{\sigma_{2}^{2} u_{2}^{2}}{2}\right)\right)
\end{split}
\end{equation*}
and this coincides with the \chf\ of the original model.
\end{note}

Starting from the explicit expression of the \chf\ one can easily compute the linear correlation coefficient at time $t$.

\begin{prop}[Correlation]
\label{prop:correlationSemeraro}
The correlation at time $t$ $\rho_{Y_{1}\left(t\right),Y_{2}\left(t\right)}$ is given by:

\begin{equation}
\rho_{Y_{1}\left(t\right)Y_{2}\left(t\right)} = \frac{\mu_{1}\mu_{2}\alpha_{1}\alpha_{2}a Var\left[H_{1}\left(t\right)\right]}{\sqrt{Var\left[Y_{1}\left(t\right)\right]Var\left[Y_{2}\left(t\right)\right]}}
\label{eqn:correlation}
\end{equation}
\end{prop}

We observe that the value of correlation $\rho$ is lower than the one obtained by \citet{Semeraro2008}. This is obvious from an intuitive point of view: in the original model the author modeled the systemic risk component using a common subordinator whilst we use two processes, $H_{1}\left(t\right),H_{2}\left(t\right)$. On the other hand, as observed before, if $a\to 1$ then $H_{1}\left(t\right)$ and $H_{2}\left(t\right)$ are indistinguishable and we retrieve the value of correlation $\rho$ obtained by \citet{Semeraro2008}.

\subsubsection{2D - Variance-Gamma}
\label{sec:2DVGSemeraro}
So far we analyzed the general model without assuming a particular form for the law of any of the processes involved. Gamma \rv's have \sd\ law then they are suitable candidates for our construction. Assuming that $\tilde{H}_{1},\tilde{H}_{2}$ has Gamma law (with a specific parameters choice) we extend Semeraro's model for the Variance Gamma process using \sd-subordinators.\\
We recall that a Gamma \rv\  has a density (\pdf) $f\left(\alpha,\beta;x\right)$ and \chf\ given by:

\begin{eqnarray}
f\left(\alpha,\beta;x\right) &=& \frac{\beta^{\alpha}}{\Gamma\left(\alpha \right)}x^{\alpha-1}e^{-\beta x}\mathbbm{1}_{x > 0}\left(x\right),
\nonumber\\
    \phi_{X}\left(u\right) &=& \left(1 - \frac{i u}{\beta}\right)^{-\alpha}
\end{eqnarray}
with $\alpha,\beta \in \mathbb{R}^{+}$. It is well-known that if $X\sim \Gamma\left(\alpha,\beta\right)$ then $cX \sim \Gamma \left(\alpha, \frac{\beta}{c}\right)$ and if $X\sim \Gamma\left(\alpha_{1},\beta\right)$ and $Y\sim \Gamma\left(\alpha_{2},\beta\right)$ are independent, then $X + Y \sim \Gamma\left(\alpha_{1} + \alpha_{2},\beta\right)$.
Now set in \eqref{eqn:subordinatedprocess}:
\begin{equation*}
I_{j} \sim \Gamma\left(A_{j},\frac{B}{\alpha_{j}}\right), \quad
H_{j} \sim \Gamma\left(A,B\right),  \quad j=1,2
\end{equation*}
and noting that $\alpha_{j}H_{j} \sim \Gamma\left(A,\frac{B}{\alpha_{j}}\right)$ we have

\begin{equation}
G_{j} \sim \Gamma\left(A_{j}+A,\frac{B}{\alpha_{j}}\right), \quad j=1,2. \nonumber
\end{equation}
Remembering that $A_{j},A,B,\alpha_{j} \in \mathbb{R}^{+}$ we have the following conditions:
\begin{align}
    & \frac{1}{A_{j} + A} = \frac{\alpha_{j}}{B}, \quad j=1,2 \label{eqn:VGfirstCondition} \\
    & 0 < \alpha_{j} \le \frac{B}{A}, \quad j=1,2 \label{eqn:VGsecondCondition}
\end{align}
Given the condition \eqref{eqn:VGfirstCondition} and \eqref{eqn:VGsecondCondition} we have that $\mathbb{E}\left[G_{j}\right] = 1$ and then $\mathbb{E}\left[G_{j}\left(t\right)\right] = t$.

\begin{note}
If we request condition \eqref{eqn:VGfirstCondition}, we have that:

\begin{equation}
1 = \alpha_{1}\frac{\left(A_{1} + A\right)}{B} = \alpha_{2}\frac{\left(A_{2} + A\right)}{B}
\nonumber
\end{equation}
and so the parameter $B$ is somehow redundant and we can assume $B=1$. 

We get the same conclusion observing that, in Equation \eqref{eqn:correlation}, we fit only the variance of $H_{1}\left(t\right)$: for this reason assuming $B=1$ is not restrictive. 

\end{note}

\noindent The following corollaries are direct application of Propositions \ref{prop:jointChfSemeraro} and  \ref{prop:correlationSemeraro}:

\begin{cor}
\label{cor:2DVGchfSemeraro}
The \chf\ in 2D Variance-Gamma case is:
\begin{align}
    \phi_{H_{j}\left(t\right)}\left(u\right) &= \left(1-i\frac{u}{B} \right)^{-t A} \quad j=1,2 \nonumber\\
    \phi_{I_{j}\left(t\right)}\left(u\right) &= \left(1-\alpha_{j}i\frac{u}{B}\right)^{-t A_{j}} \quad j=1,2\nonumber \\
    \phi_{Z_{a}\left(t\right)}\left(u \right) & = \frac{\phi_{H_{1}\left(t\right)}\left(u\right)}{\phi_{H_{1}\left(t\right)}\left(a u\right)} = \left(\frac{B -i u}{B-i a u}\right)^{-t A}
\end{align}
and so \chf\ $\phi_{\boldsymbol{Y}\left(t\right)}\left(\boldsymbol{u}\right)$ in \eqref{eqn:generalchf} can be computed.
\end{cor}

\begin{cor}
Linear correlation coefficient  in 2D Variance-Gamma case is given by:
\begin{equation*}
\rho_{\left(Y_{1}\left(t\right),Y_{2}\left(t\right)\right)} = \frac{\mu_{1}\mu_{2}\alpha_{1}\alpha_{2}a A}{\sqrt{\sigma_{1}^{2} + \mu_{1}^{2}\alpha_{1}}\sqrt{\sigma_{2}^{2} + \mu_{2}^{2}\alpha_{2}}}
\end{equation*}
\end{cor}

\subsection{Extension of Semeraro-Luciano's Model}
\label{sec:LSSD}
The model presented by \citet{SL2010}, which was developed in order to catch those correlation levels in log-returns that the model proposed by \citet{Semeraro2008} is not able to get (see \citet{WD2010}), can be extended in a similar way to what we showed in Section \ref{sec:SSD}.
\begin{defn}[\sd-Luciano and Semeraro's model]
Let $I_{j}\left(t\right), \; j=1,2$, subordinators and let $H_{1}\left(t\right)$ and $H_{2}\left(t\right)$ two \sd\ subordinators independent from $I_{j}\left(t\right)$. Define the following process:


\begin{equation}
\boldsymbol{Y}^{\rho}\left(t\right) =
				\left(
				\begin{array}{ll}
				\mu_{1} I_{1}\left(t\right) + \sigma_{1}W_{1}\left(I_{1}\left(t\right)\right) +\alpha_{1}\mu_{1}H_{1}\left(t\right) + \sqrt{\alpha_{1}}\sigma_{1}W_{1}^{\rho}\left(H_{1}\left(t\right)\right) \\
\mu_{2} I_{2}\left(t\right) + \sigma_{2}W_{2}\left(I_{2}\left(t\right)\right) +\alpha_{2}\mu_{2}H_{2}\left(t\right) + \sqrt{\alpha_{2}}\sigma_{2}\left(W_{2}^{\rho}\left(aH_{1}\left(t\right)\right) + \tilde{W}\left(Z_{a}\left(t\right)\right)\right)
				
 				\end{array}\right)
				\label{eqn:LSgeneralization}
\end{equation}
where $W_{1}\left(t\right)$ and $W_{2}\left(t\right)$ are independent \BM's whereas
\begin{equation*}
\mathbb{E}\left[dW_{1}^{\rho}\left(t\right)dW_{2}^{\rho}\left(t\right)\right] = \rho dt
\end{equation*}
and $\tilde{W}\left(t\right)$ is independent from $\boldsymbol{W}\left(t\right) = \left(W_{1}\left(t\right),W_{2}\left(t\right)\right)$ and $\boldsymbol{W}^{\rho}\left(t\right) = \left(W_{1}^{\rho}\left(t\right),W_{2}^{\rho}\left(t\right)\right)$.
\end{defn}

Here too the \chf\ has a nice closed expression. 
\begin{prop}[Characteristic Function]
\label{prop:jointChfSemeraroLuciano}
The joint \chf\ $\phi_{\boldsymbol{Y}^{\rho}\left(t\right)}\left(\boldsymbol{u}\right)$ of the process $\boldsymbol{Y}^{\rho}\left(t\right) = \left(Y_{1}^{\rho}\left(t\right),Y_{2}^{\rho}\left(t\right)\right)$ at time $t$ defined in \eqref{eqn:LSgeneralization} is given by:

\begin{equation}
\begin{split}
\phi_{\boldsymbol{Y}\left(t\right)^{\rho}}\left(\boldsymbol{u}\right) = & \phi_{I_{1}\left(t\right)}\left(u_{1}\mu_{1} + \frac{i}{2}\sigma_{1}^{2}u_{1}^{2}\right) \phi_{I_{2}\left(t\right)}\left(u_{2}\mu_{2} + \frac{i}{2}\sigma_{2}^{2}u_{2}^{2}\right)\\
 & \phi_{H_{1}\left(t\right)}\left(\frac{i}{2}u_{1}^{2}\alpha_{1}\sigma_{1}^{2}\left(1-a\right) +  \boldsymbol{u}^{T}\boldsymbol{\mu} +\frac{i}{2}\boldsymbol{u}^{T}a\Sigma \boldsymbol{u}\right) \phi_{Z_{a}\left(t\right)}\left(u_{2}\mu_{2}\alpha_{2} + \frac{i}{2}u_{2}^{2}\alpha_{2}\sigma_{2}^{2}\right)
\end{split}
\nonumber
\end{equation}
where $\boldsymbol{\mu} = \left[\alpha_{1}\mu_{1},a\alpha_{2}\mu_{2}\right]$ and

\[\Sigma =
\begin{bmatrix}
	\alpha_{1}\sigma_{1}^{2} & \sqrt{\alpha_{1}\alpha_{2}}\sigma_{1}\sigma_{2}\rho \\
	\sqrt{\alpha_{1}\alpha_{2}}\sigma_{1}\sigma_{2}\rho & \alpha_{2}\sigma_{2}^{2}
\end{bmatrix}
\]
\end{prop}

Following the technique proposed for the proof of Proposition \ref{prop:correlationSemeraro} one can show the following:

\begin{prop}[Correlation]
The correlation at time $t$, $\rho_{Y_{1}^{\rho}\left(t\right),Y_{2}^{\rho}\left(t\right)}$ is given by:
\begin{equation}
\rho_{Y_{1}^{\rho}\left(t\right),Y_{2}^{\rho}\left(t\right)} = \frac{a\left(\mu_{1}\mu_{2}\alpha_{1}\alpha_{2}Var\left[H_{1}\left(t\right)\right] +  \rho\sigma_{1}\sigma_{2}\sqrt{\alpha_{1}\alpha_{2}}\E\left[H_{1}\left(t\right)\right]\right)}{\sqrt{Var\left[Y_{1}\left(t\right)\right]Var\left[Y_{2}\left(t\right)\right]}}
\end{equation}
\end{prop}
\vspace{0.3cm}
All considerations about correlation coefficient and \chf\ we pointed out in Section \ref{sec:SSD} are still valid.

\subsubsection{2D - Variance-Gamma}
Here too it's possible to build a 2D-Variance Gamma process by choosing
\begin{equation*}
I_{j} \sim \Gamma \left(A_{j},\frac{B}{\alpha_{j}}\right), \;
H_{j} \sim \Gamma\left(A,B\right),\; j=1,2
\end{equation*}
We have that:

\begin{equation*}
I_{j} + \alpha_{j}H_{j} \sim \Gamma \left(A_{j} + A, \frac{B}{\alpha_{j}}\right) , \; j=1,2
\end{equation*}
and, imposing conditions \eqref{eqn:VGfirstCondition} and \eqref{eqn:VGsecondCondition}, we have get $\E\left[G_{j}\right] = 1$ and, consequently, $\E\left[G_{j}\left(t\right)\right] = t$ for $j=1,2$.
Following the same argument of Section \ref{sec:2DVGSemeraro}, expressions of linear correlation coefficient and the \chf\ for the 2D Variance Gamma case can be derived.

\begin{cor}
\label{cor:linearcoefunSL}
	Linear correlation coefficient  in 2D Variance-Gamma case is given by:
	\begin{equation*}
	\rho_{\left(Y_{1}^{\rho}\left(t\right),Y_{2}^{\rho}\left(t\right)\right)} = \frac{a\left(\mu_{1}\mu_{2}\alpha_{1}\alpha_{2} A +\rho A \sigma_{1}\sigma_{2}\sqrt{\alpha_{1}\alpha_{2}}\right)}{\sqrt{\sigma_{1}^{2} + \mu_{1}^{2}\alpha_{1}}\sqrt{\sigma_{2}^{2} + \mu_{2}^{2}\alpha_{2}}}
	\end{equation*}
\end{cor}

The \chf\ can be obtained combining Corollary \ref{cor:2DVGchfSemeraro} with Proposition \ref{prop:jointChfSemeraroLuciano}.

\subsection{Extension of Ballotta-Bonfiglioli's Model}
\label{sec:BBSD}
The construction technique of dependent \Levy\ processes proposed by \citet{BB2013} is slightly different from what we have seen so far. The dependence between processes is not introduced on subordinators, as in the previous case, but two subordinated \BM of the same type are added together. Some convolution conditions on parameters guarantee that the resulting process is of the same type of the summed ones. This model, as the previous ones, can be extended using \sd\ subordinators.

\begin{defn}[\sd-Ballotta and Bonfiglioli's model]
\label{def:BBmodelsd}
Let $H_{1}\left(t\right)$ and $H_{2}\left(t\right)$ be \sd\ subordinators as in \eqref{eqn:subordinatorsSD} and define:

\begin{equation}
\boldsymbol{Y}\left(t\right) = \left(Y_{1}\left(t\right),Y_{2}\left(t\right)\right) = \left(X_{1}\left(t\right) + a_{1}R_{1}\left(t\right), X_{2}\left(t\right) + a_{2} R_{2}\left(t\right)\right)
\label{eqn:extendedModelBB}
\end{equation}
where:

\begin{itemize}
\item $X_{j}\left(t\right)$ is a subordinated Brownian motion with parameters $\left(\beta_{j},\gamma_{j},\nu_{j}\right), \; j=1,2$, where $\beta_{j} \in \mathbb{R}$ is the drift, $\gamma_{j}\in \mathbb{R}^{+}$ is the diffusion and $\nu_{j} \in \mathbb{R}^{+}$ is the variance of the subordinator. We state the two independent subordinators of $X_{j}\left(t\right)$ with $G_{j}\left(t\right)$. We have that:
\begin{equation*}
X_{j}\left(t\right) = \beta_{j}G_{j}\left(t\right) + W_{j}\left(G_{j}\left(t\right)\right),\quad j=1,2.
\end{equation*}
\item $R_{1}\left(t\right) $ and $R_{2}\left(t\right)$ are given by:
\begin{align}
R_{1}\left(t\right) & = \beta_{R_{1}}H_{1}\left(t\right) + \gamma_{R_{1}}W\left(H_{1}\left(t\right)\right) \nonumber \\
R_{2}\left(t\right) & = \beta_{R_{2}}H_{2}\left(t\right) + \gamma_{R_{2}}\left(W\left(aH_{1}\left(t\right)\right) + \tilde{W}\left(Z_{a}\left(t\right)\right)\right) \label{eqn:Rprocess}
\end{align}
where $W\left(t\right)$ and $\tilde{W}\left(t\right)$ are independent Brownian motions and $\beta_{R_{j}}\in \mathbb{R}$ and $\gamma_{R_{j}} \in \mathbb{R}^{+}$.
\end{itemize}
\end{defn}

The following Lemma will help to derive the \chf\ of the process.
\begin{lem}
\label{lem:01}
The \chf\ of the process defined in \eqref{eqn:Rprocess} at time $t$ is given by:

\begin{equation*}
\begin{split}
\phi_{\boldsymbol{R}\left(t\right)}\left(\boldsymbol{u}\right) =& \phi_{H_{1}\left(t\right)}\left(
u_{1}\beta_{R_{1}} + u_{2}\beta_{R_{2}}a + \frac{i}{2}\left(u_{1}^{2}\gamma_{R_{1}}^{2} +2u_{1}u_{2}\gamma_{R_{1}}\gamma_{R_{2}}a +  u_{2}^{2}a\gamma_{R_{2}}^{2}\right)\right) \\
&  \phi_{Z_{a}\left(t\right)}\left(u_{2}\beta_{R_{2}} +\frac{i}{2}u_{2}^{2}\gamma_{R_{2}}^{2}\right)
\end{split}
\end{equation*}
\end{lem}

The \chf\ of the process defined in \eqref{eqn:extendedModelBB} is given by the following Proposition.

\begin{prop}[Characteristic Function]
\label{lem:01bis}
The \chf\ of the process at time $t$ defined in \eqref{eqn:extendedModelBB} is given by:

\begin{equation}
\begin{split}
\phi_{\boldsymbol{Y}\left(t\right)}\left(u_{1},u_{2}\right) = & \phi_{G_{1}\left(t\right)}\left(\beta_{1}u_{1} + \frac{i}{2}u_{1}^{2}\gamma_{1}^{2}\right) \\
& \phi_{G_{2}\left(t\right)}\left(\beta_{2}u_{2} + \frac{i}{2}u_{2}^{2}\gamma_{2}^{2}\right) \\
& \phi_{\boldsymbol{R}\left(t\right)}\left(\boldsymbol{a}\circ\boldsymbol{u}\right)
\label{eqn:BBExtended}
\end{split}
\end{equation}
where $\boldsymbol{a} = \left(a_{1},a_{2}\right)$ and $\boldsymbol{u} = \left(u_{1},u_{2}\right)$ and $\circ$ is the Hadamard product.

\end{prop}

\begin{note}
As in the precious models it is easy to verify that:
\begin{equation}
\begin{split}
\lim\limits_{\substack{a\to 1 \\
 \beta_{R_{1}},\beta_{R_{2}} \to \beta_{Z} \\
\gamma_{R_{1}},\gamma_{R_{2}} \to \gamma_{Z}}} & \phi_{\boldsymbol{Y}\left(t\right)}\left(u_{1},u_{2}\right)  =  \phi_{G_{1}\left(t\right)}\left(\beta_{1}u_{1} + \frac{i}{2}u_{1}^{2}\gamma_{1}^{2}\right) \\
& \phi_{G_{2}\left(t\right)}\left(\beta_{2}u_{2} + \frac{i}{2}u_{2}^{2}\gamma_{2}^{2}\right) \phi_{Z\left(t\right)}\left(\beta_{Z}\left(a_{1}u_{1}+a_{2}u_{2}\right) + \frac{i}{2}\left(a_{1}u_{1} + a_{2}u_{2}\right)^{2}\gamma_{Z}^{2}\right)
\end{split}
\nonumber
\end{equation}
which is the \chf\ obtained by \citet{BB2013}.
\end{note}
Even then, the correlation coefficient of the process $\boldsymbol{Y}\left(t\right)$ can be obtained.


\begin{prop}
\label{prop:CorrelationBB}
The correlation coefficient at time $t$ of the process $\boldsymbol{Y}\left(t\right)$ defined in \eqref{eqn:extendedModelBB} is given by:

\begin{equation}
\rho_{\boldsymbol{Y}\left(t\right)} = \frac{a_{1}a_{2}a\left(\beta_{R_{1}}\beta_{R_{2}} Var\left[H_{1}\left(t\right)\right] + \gamma_{R_{1}}\gamma_{R_{2}}\mathbb{E}\left[H_{1}\left(t\right)\right]\right)
}{\sqrt{Var\left[Y_{1}\left(t\right)\right]}\sqrt{Var\left[Y_{2}\left(t\right)\right]}}
\end{equation}
\end{prop}

\subsubsection{Convolution Conditions}
It's easy to show that, if $X_{j}\left(t\right)$ and $R_{j}\left(t\right),\; j=1,2$, are subordinated \BM's with subordinators from the same family, then $Y_{j}\left(t\right)$ is a subordinated process of the same type of $X_{j}\left(t\right)$ and $R_{j}\left(t\right)$ if the following \citet{BB2013} style convolution conditions hold:
\begin{equation}
\nu_{R} \coloneqq \nu_{R_{1}} = \nu_{R_{2}}
\label{eqn:BBSubordinatorsVariance}
\end{equation}
and
\begin{equation}
\left\{
\begin{array}{l}
\alpha_{j}\mu_{j} = \nu_{R}a_{j}\beta_{R_{j}} \quad j=1,2 \\
\alpha_{j}\sigma_{j}^{2} = \nu_{R}a_{j}^{2}\gamma_{R_{j}}^2 \quad j=1,2
\end{array}
\right.
\label{eqn:convconditions}
\end{equation}
Relation \eqref{eqn:BBSubordinatorsVariance} holds because $H_{1}\left(t\right)$ and $H_{2}\left(t\right)$ have the same law and so they have the same variance $\nu_{R}$. It is easy to check that if Equations \eqref{eqn:convconditions} are satisfied then:
\begin{equation}
\mu_{j} = \beta_{j} + a_{j}\beta_{R_{j}}, \quad \sigma_{j}^2 = \gamma_{j}^2 + a_{j}^2\gamma_{R_{j}}^2, \quad
\alpha_{j} = \nu_{j}\nu_{R}/\left(\nu_{j} + \nu_{R}\right).
\nonumber
\end{equation}

\subsubsection{2D - Variance-Gamma}
We can construct a 2D -  Variance-Gamma using Gamma subordinators as follows.
\begin{itemize}
\item Let $H_{1}\left(t\right) \sim \Gamma\left(\frac{t}{\nu_{R}},\frac{1}{\nu_{R}}\right)$ be a Gamma subordinator and set $H_{2}\left(t\right) = aH_{1}\left(t\right) + Z_{a}\left(t\right)$.
\item Let $R_{j}\left(t\right)$ be a subordinated \BM\ (with drift $\beta_{R_{j}}$ and diffusion $\gamma_{R_{j}}$) obtained using the Gamma subordinator $H_{j}\left(t\right) \sim \Gamma\left(\frac{t}{\nu_{R}},\frac{1}{\nu_{R}}\right), \; j=1,2$.
\item Let $X_{j}\left(t\right)$ be a subordinated \BM\ (with drift $\beta_{j}$ and diffusion $\gamma_{j}$) obtained using a Gamma subordinator $G_{j}\left(t\right) \sim \Gamma\left(\frac{t}{\nu_{j}},\frac{1}{\nu_{j}}\right), \; j=1,2$.
\item Set $Y_{j}\left(t\right) = X_{j}\left(t\right) + a_{j}R_{j}\left(t\right)$
\end{itemize}

\noindent We obtain that $Y\left(t\right) \sim VG\left(\mu_{j},\sigma_{j},\alpha_{j}\right)$, $j=1,2$, where $\mu_{j},\sigma_{j},\alpha_{j}$ respect convolution conditions \eqref{eqn:convconditions}.

\vspace{0.5cm}
The joint \chf\ $\phi_{\boldsymbol{Y}\left(t\right)}\left(u_{1},u_{2}\right)$ can be easly derived using \eqref{eqn:BBExtended} and remembering the expression of the \chf\ of a $\Gamma\left(\alpha,\beta\right)$ \rv :

\begin{equation*}
\phi\left(u\right) = \left(1 - \frac{iu}{\beta}\right)^{-\alpha}
\end{equation*}

Applying Proposition \ref{prop:CorrelationBB} one can derive the correlation coefficient of the 2D - Variance-Gamma process which has the following expression:

\begin{equation*}
\rho_{\boldsymbol{Y}\left(t\right)} = \frac{a_{1}a_{2}a\left(\beta_{R_{1}}\beta_{R_{2}}\nu_{R}+ \gamma_{R_{1}}\gamma_{R_{2}}\right) }{\sqrt{\sigma_{1}^{2} + \mu_{1}^{2}\alpha_{1}}\sqrt{\sigma_{2}^{2} + \mu_{2}^{2}\alpha_{2}}}
\end{equation*}





\section{Financial Application}
\label{sec:NumRes}
So far we derived the theoretical modeling framework and we showed how to build correlated \Levy\ processes using \sd\ subordinators. In this section we show a real application of models presented in Section \ref{sec:modExtension} to energy  markets. Many standard techniques for market modeling, calibration, paths simulation and pricing can be adapted to our case. \\

Similar to what already done in  \citet{CT2003}, we model energy forward markets by defining exponential \Levy\ processes using the process $\boldsymbol{Y}\left(t\right)$ derived in Section \ref{sec:modExtension}. The forward price $F_{j}(t),\; j=1,2$ at time $t$ can be defined as follow:
\begin{equation}
F_{j}\left(t\right) = F_{j}\left(0\right) e^{\omega_{j} t + Y_{j}\left(t\right) }
\label{eqn:themodel}
\end{equation}
where $\omega_{j}$ is the drift correction that leads us to work under a risk-neutral probability measure. 
Non-arbitrage conditions can be obtained setting:
\begin{equation}
\omega_{j} = -\varphi_{j}\left(-i\right)
\label{eqn:omegaexpr}
\end{equation}
where $\varphi_{j}\left(u\right)$ is the characteristic exponent of the process $Y_{j}\left(t\right)$.\\

In order to calibrate our model we use a two steps calibration procedure, as the one proposed by \citet{SL2010}. It is worthwhile noticing that marginal distributions don't depend on the parameters we use to model dependence structures. Then, if we observe in the market $n$ quoted vanilla products $\left(C_{i}\right)_{i=1}^{n}$ we can obtain the marginal parameter vector $\boldsymbol{\theta}^{*}$ solving the following:
\begin{equation}
\boldsymbol{\theta}^{*} = \argmin_{\boldsymbol{\theta}} \sum_{i=1}^{n} \left(C_{i}^{\boldsymbol{\theta}}\left(K,T\right) - C_{i}\right)^{2}.
\label{eqn:minimizationproblem_margins}
\end{equation}
where $C_{i}^{\boldsymbol{\theta}}\left(K,T\right)$ are model prices.

Once we fit $\boldsymbol{\theta}^{*}$ we have to calibrate dependence structure. Generally derivatives written on multiple underlying assets are not very liquid: for this reason the dependence parameters vector $\boldsymbol{\eta}^{*}$ is estimated fitting the correlation matrix on historical data. Theoretical correlation matrix can be computed using the closed form expression for linear correlation coefficients derived in Section \ref{sec:modExtension}.
In the first step we used a NLLS approach combined with the FFT method proposed by \citet{CarrMadan99} (the version proposed by \citet{lewis} leads to similar results), whereas in the second step both NLLS and GMM method can be used: in our experiments we adopted the first one.\\

An observant reader would point out that 2D-Variance Gamma processes can be easily simulated by using standard techniques presented, for example, in \citet{Dev86} and \citet{CT2003}. The only arising difficulty is the simulation of $Z_{a}\left(t\right)$ processes. \citet{cs20, cs20_2} have shown that the \arem\ $Z_{a}$ of a Gamma distribution $\Gamma\left(\alpha,\lambda\right)$ can be exactly simulated by taking:
\begin{equation*}
Z_{a} = \sum_{j=1}^{S} X_{j}
\end{equation*}
where

\begin{equation*}
S \sim \mathfrak{B}\left(\alpha,1-a\right) \quad X_{j} \sim \mathfrak{E}\left(\lambda/a\right) \quad X_{0} = 0 \quad \mathbb{P}-a.s.
\end{equation*}
$\mathfrak{B}\left(\alpha,1-a\right)$ denote a Polya or negative binomial distribution and $\mathfrak{E}\left(\lambda/a\right)$ denotes an exponential distribution. Using this result a simulation scheme can be derived and a Monte Carlo algorithm for pricing purposing developed.\\

One can argue that, alternatively to MC schemes, since the \chf's of the log-process are known in closed form, Fourier methods can be adopted. Different techniques based on Fourier Transform are available for pricing, and some of them can be used in a multivariate contest (see for example \citet{hurd2009}, \citet{Pellegrino2016} and \citet{caldanafusai2016}). In this section we used the method proposed by \citet{caldanafusai2016} which gives a good approximation for spread-options prices and it's simpler to implement than the one proposed by \citet{hurd2009}, because it requires only one Fourier inversion.\\

\begin{figure}
    \centering
    \includegraphics[scale=0.25]{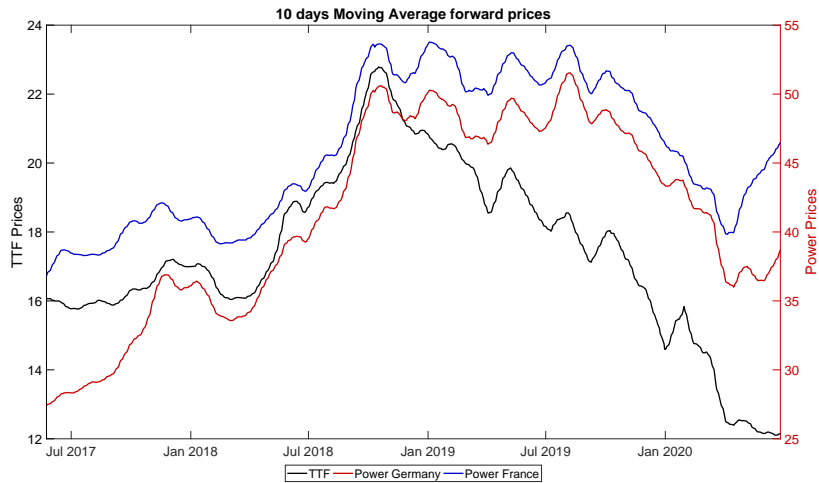}
    \caption{German, Franch and Naural Gas TTF forward market.}
    \label{fig:RealMarketPath}
\end{figure}

The remaining part of the section is split into two branches:
in the first one we apply our models to the German and French power forward markets, whereas in the second part we focus on German power forward market and to natural gas forward market.
We have chosen those markets because, in the first case we deal with markets that are strongly correlated due to the configuration of European electricity network, whereas in the second case, the correlation between markets is still positive, since natural gas can be used to produce electricity, but it's not as strong as in the former case. This gives us the opportunity to test our models for different level of correlations. \\
Moreover, as can be observed by Figure \ref{fig:RealMarketPath}, due to the structure of European electricity grid, power markets usually react \enquote{in the same way at the same time} whereas a delay between power markets and natural gas market is more likely. Then we expect a value for the parameter $a$ very close to one between forward power markets and a lower value when we consider forward power and natural gas markets.

\vspace{0.4cm}
For the sake of concision we use the following notation:

\begin{itemize}
\item (\emph{SSD}): \sd-Semeraro's model presented in Section \ref{sec:SSD}.
\item (\emph{LSSD}): \sd-Luciano and Semeraro's model presented in Section \ref{sec:LSSD}.
\item (\emph{BBSD}): \sd-Ballotta and Bonfiglioli's model presented in Section \ref{sec:BBSD}.
\end{itemize}
\noindent In our experiments we price spread options on future prices, denoted $F_{i}\left(t\right), i=1,2$, whose payoff is given by:

\begin{equation*}
\Phi_{T} = \left(F_{1}\left(T\right) - F_{2}\left(T\right) - K\right)^{+}.
\end{equation*}
It customary to reserve the name  \textit{Cross-Border} or \textit{Spark-Spread}  option if the futures are relative to power or gas markets, respectively.
In all our experiments we use the MC technique with $N_{sim}=10^{6}$ simulations and the Fourier-based method proposed by \citet{caldanafusai2016}.


\subsection{Application to German and French Power Markets}
In order to calibrate our model we need both derivatives contracts written on forward and historical time series of forward quotations. The data-set\footnote{Data Source: www.eex.com.} we relied upon is composed as follow:
\begin{itemize}
\item Forward quotations from 25 April 2017 to 12 November 2018 of Calendar 2019 power forward. A Forward Calendar 2019 contract is a contract between two counterparts to buy or sell a specific volume of energy in MWh at fixed price for all the hours of 2019. Calendar power forward in German and France are stated respectively with DEBY and F7BY.
\item Call Options on power forward 2019 quotations for both countries with settlement date 12 November 2018. We used strikes in a range of $\pm 10\, [EUR/MWh]$ around the settlement price of the Forward contract, i.e. we exclude deep ITM and OTM options.
\item We assume risk-free rate $r=0.015$.
\item The historical correlation between markets is $\rho_{mkt} = 0.94$.
\end{itemize}

\begin{figure}
    \centering
    \includegraphics[scale=0.25]{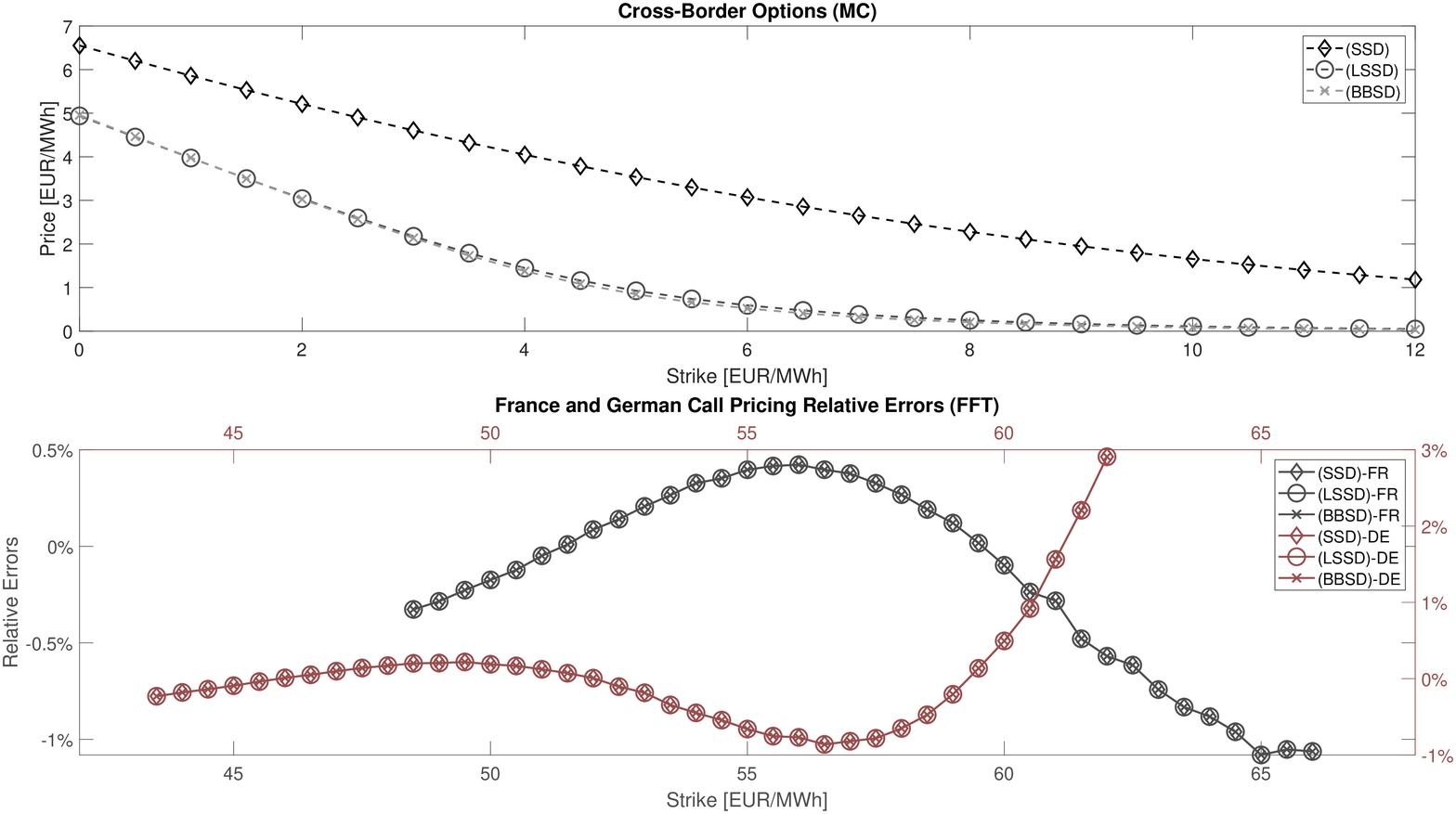}
    \caption{Percentage errors and Cross Border option prices.}
    \label{fig:CrossBorderOption}
\end{figure}

From Table \ref{tbl:commoparams} we see that all models provide the same set of marginal parameters. In the lower box of Figure \ref{fig:CrossBorderOption} we report the percentage error $\epsilon_{i}$ defined as:
\begin{equation*}
\epsilon_{i} = \frac{C_{i}^{\theta}\left(K,T\right) - C_{i}}{C_{i}}.
\end{equation*}
We can observe this error is really small, varying $K$: our model is able to replicate market prices and therefore can be used for pricing purposes.  

If we look at the fitted correlation the situation is slightly different. The \emph{SSD} model presented in Section \ref{sec:SSD} fits a correlation that is roughly zero. For this reason the model is not recommendable for \textit{Cross-Border} option pricing because it overestimates the derivative price as we can see from the upper picture in Figure \ref{fig:CrossBorderOption}. The \emph{LSSD} model of Section \ref{sec:LSSD} is better than the previous one and the fitted correlation is very close to the one observed in the market as we can see from Table \ref{table:depparamsLSSDCB}. For this reason the \emph{LSSD} model can be used to price \textit{Cross-Border} options. The \emph{BBSD} model derived by in Section \ref{sec:BBSD} provides an even better fitting of market correlation. We conclude that the  \emph{BBSD} model is the best one for the valuation of \textit{Cross-Border} options. A comparison between models can be found in the upper part of Figure \ref{fig:CrossBorderOption}: option prices provided by the \emph{BBSD} model are the lowest ones due to the highest value of fitted correlation.
\par One additional consideration is needed: we note that, as we expected, for all models, the fitted value for the \sd\ parameters $a$ is very close to one. This is not a surprise because German and France forward markets are so strictly correlated that whenever an event occurs in a market it has an immediate impact on the other one. As mentioned before, if $a \to 1$ we obtain the original models of \citet{Semeraro2008}, \citet{SL2010} and \citet{BB2013}. For this reason, for \emph{Cross Border} options, there's not an essential difference between original models and the extended ones.

\begin{table}[!htb]
\scriptsize

\begin{minipage}{1\linewidth}
\centering
    \begin{tabular}[t]{ccccccc}
    \toprule
    Model & $\mu_{1}$ & $\mu_{2}$ & $\sigma_{1}$ & $\sigma_{2}$ & $\alpha_{1}$ & $\alpha_{2}$  \\ [0.5ex]
    \midrule
\emph{SSD} & 0.40 & 0.61  & 0.31 & 0.32 & 0.02 & 0.02   \\
\emph{LSSD} & 0.40 & 0.61  & 0.31 & 0.32 & 0.02 & 0.02   \\
\emph{BBSD} & 0.40 & 0.61  & 0.31 & 0.32 & 0.02 & 0.02   \\
\bottomrule
\end{tabular}
\caption{Fitted marginal parameters for German and French power markets.}
\label{tbl:commoparams}
\end{minipage}\hfill
\vspace{0.2cm}
\begin{minipage}{.23\linewidth}
\centering

\begin{tabular}[t]{cc}
\toprule
Parameter & Value \\ [0.5ex]
\midrule
$A$ & 41.89  \\
$B$ & 1.00  \\
$a$ & 0.99  \\
$\rho_{mod}$ & 0.05  \\ [1ex]
\bottomrule\end{tabular}
\caption{\emph{SSD}}
\label{table:depparamsSSDCB}
\end{minipage}\hfill
\begin{minipage}{.3\linewidth}
\centering

\smallskip

\begin{tabular}[t]{cc}
\toprule
Parameter & Value \\ [0.5ex]
\midrule
$A$ & 42.31  \\
$B$ & 1.00  \\
$\rho$ & 1.00  \\
$a$ & 0.99  \\
$\rho_{mod}$ & 0.92  \\ [1ex]
\bottomrule
\end{tabular}
\caption{\emph{LSSD}}
\label{table:depparamsLSSDCB}
\end{minipage}\hfill
\begin{minipage}{.4\linewidth}
\centering

\smallskip

\begin{tabular}[t]{cccc}
\toprule
Parameter & Value & Parameter & Value\\ [0.5ex]
\midrule
$\beta_{1}$ & -0.00 & $\beta_{R_{2}}$ & 0.85 \\
$\beta_{2}$ & 0.09  & $\gamma_{R_{1}}$ & 0.50\\
$\gamma_{1}$ & 0.00 & $\gamma_{R_{2}}$ & 0.47  \\
$\gamma_{2}$ & 0.10 & $\nu_{R}$ & 0.02  \\
$\nu_{1}$ & 1.01 & $a$ & 0.99  \\
$\nu_{2}$ & 0.14 & $\rho_{mod}$ & 0.94  \\
$\beta_{R_{1}}$ & 0.62 & &\\ [1ex]
\bottomrule
\end{tabular}
\caption{\emph{BBSD}}
\label{table:depparamsBBSDCB}
\end{minipage}
\end{table}

\subsection{Application to Power German and TTF Gas Future Market}
\label{sec:VGtoSparkSpread}
In this section we apply our models to the German power forward market and to the Natural Gas forward market (TTF). These two markets are positively correlated but not as strongly as power markets are. \\
As in the power case, data-set\footnote{Data Source: www.eex.com and www.theice.com} we relied upon is the following one:

\begin{itemize}
\item Forward quotations from 1 July 2019 to 09 September 2019 relative to the Month January 2019 for the Power Forward in Germany and  the Gas TTF Forward.
\item Call Options on power forward January 2020 quotations for both Germany and TTF with settlement date 9 September 2019. As done before, we use strikes prices $K$ in a range of $\pm 10\, [EUR/MWh]$ around the settlement price of the forward contract, i.e. we exclude deep ITM and OTM options.
\item We assume risk-free rate $r=0.015$.
\item The historical correlation between log-returns is $\rho_{mkt} = 0.54$.
\end{itemize}

\begin{figure}
    \centering
    \includegraphics[scale=0.25]{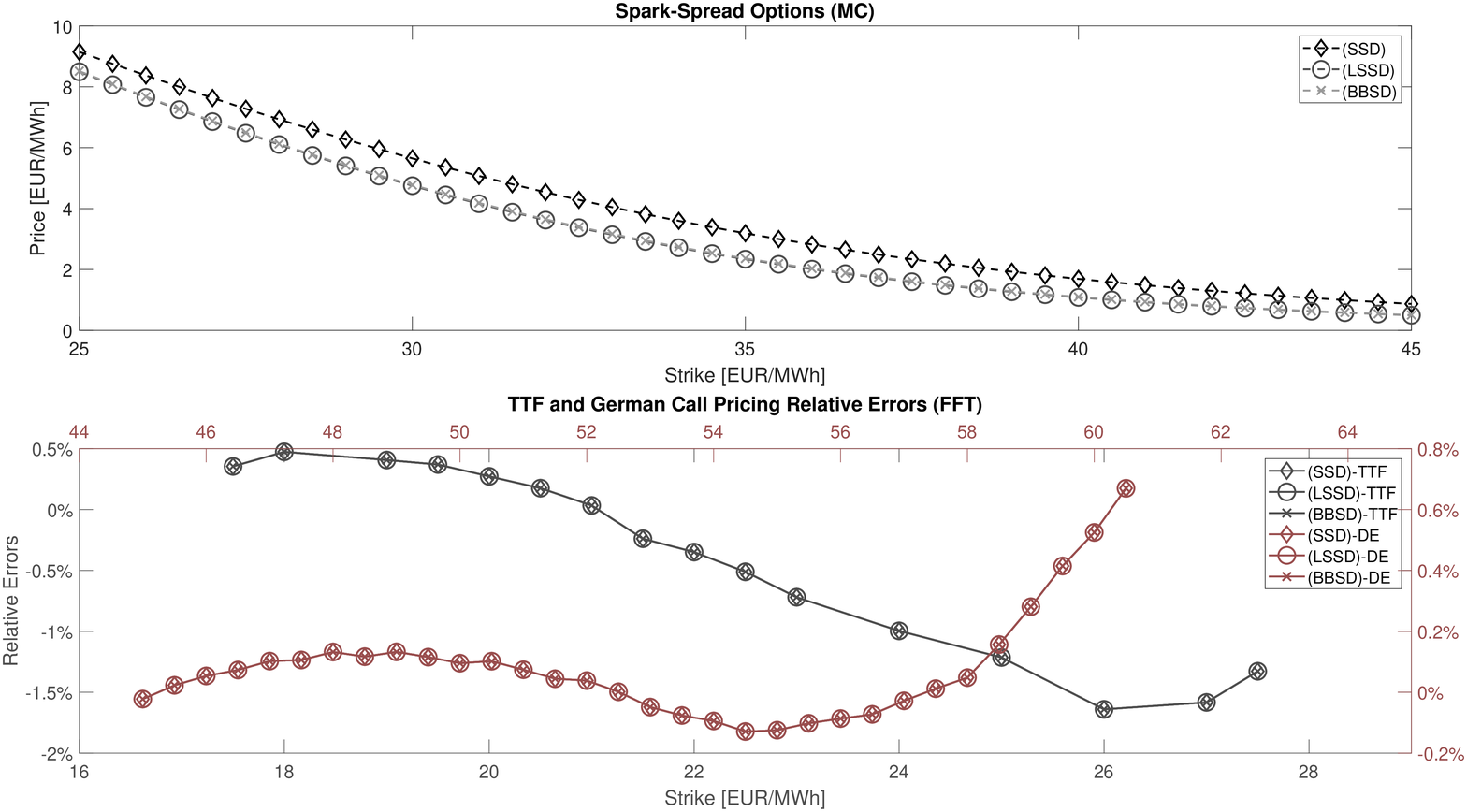}
    \caption{Percentage errors and Spark-Spread option prices.}
    \label{fig:SparkSpreadOption}
\end{figure}

In the picture at the bottom of Figure \ref{fig:SparkSpreadOption} we can see that all models provide a good fitting of quoted market options because the error $\epsilon$ is very small. In Figure \ref{fig:SparkSpreadOption} the picture at the top shows that the \emph{SSD} model overprices the \textit{Spark-Spread} option due to the fact that captured correlation is close to zero. Both \emph{LSSD} and \emph{BBSD} models provide a lower price of the derivatives because they are able to catch the market correlation.
Fitted parameters are shown in Table \ref{tbl:commoparams_SS}: we observe that the \sd\ parameter $a$ is no more as close to one as it was in the forward power markets. This result is reasonable for different reasons. First of all only approximately the 25\% of electricity in Germany is produce using natural gas: for this reason if natural gas prices falls the effect on electricity prices could not be immediate. Moreover, despite of what happens for electricity, natural gas can be stored. Many electricity producers subscribe swing contracts to protect against perturbations in natural gas prices. Then a sudden but temporary change in gas market prices doesn't effect the cost of producing electricity and consequently its price. Of course if the perturbation in natural gas prices last too long, after a while one should expect to observe the perturbation in electricity prices too.   
\begin{table}[!htb]
\scriptsize

\begin{minipage}{1\linewidth}
\centering
    \begin{tabular}[t]{ccccccc}
    \toprule
    Model & $\mu_{1}$ & $\mu_{2}$ & $\sigma_{1}$ & $\sigma_{2}$ & $\alpha_{1}$ & $\alpha_{2}$  \\ [0.5ex]
    \midrule
\emph{SSD} & 0.46 & 0.24  & 0.43 & 0.33 & 0.08 & 0.05   \\
\emph{LSSD} & 0.46 & 0.24  & 0.43 & 0.33 & 0.08 & 0.05   \\
\emph{BBSD} & 0.46 & 0.24  & 0.43 & 0.33 & 0.08 & 0.05   \\
\bottomrule
\end{tabular}
\caption{Fitted marginal parameters for power and gas forward markets.}
\label{tbl:commoparams_SS}
\end{minipage}\hfill
\vspace{0.2cm}
\begin{minipage}{.23\linewidth}
\centering

\begin{tabular}[t]{cc}
\toprule
Parameter & Value \\ [0.5ex]
\midrule
$A$ & 12.36  \\
$B$ & 1.00  \\
$a$ & 0.99  \\
$\rho_{mod}$ & 0.04  \\ [1ex]
\bottomrule\end{tabular}
\caption{\emph{SSD}}
\label{table:depparamsSSDSS}
\end{minipage}\hfill
\begin{minipage}{.3\linewidth}
\centering

\smallskip

\begin{tabular}[t]{cc}
\toprule
Parameter & Value \\ [0.5ex]
\midrule
$A$ & 9.89  \\
$B$ & 1.00  \\
$\rho$ & 0.89  \\
$a$ & 0.90  \\
$\rho_{mod}$ & 0.57  \\ [1ex]
\bottomrule
\end{tabular}
\caption{\emph{LSSD}}
\label{table:depparamsLSSDSS}
\end{minipage}\hfill
\begin{minipage}{.4\linewidth}
\centering

\smallskip

\begin{tabular}[t]{cccc}
\toprule
Parameter & Value & Parameter & Value\\ [0.5ex]
\midrule
$\beta_{1}$ & 0.13 & $\beta_{R_{2}}$ & 0.29 \\
$\beta_{2}$ & 0.12  & $\gamma_{R_{1}}$ & 0.47\\
$\gamma_{1}$ & 0.23 & $\gamma_{R_{2}}$ & 0.29  \\
$\gamma_{2}$ & 0.23 & $\nu_{R}$ & 0.11  \\
$\nu_{1}$ & 0.28 & $a$ & 0.90  \\
$\nu_{2}$ & 0.12 & $\rho_{mod}$ & 0.54  \\
$\beta_{R_{1}}$ & 0.47 & &\\ [1ex]
\bottomrule
\end{tabular}
\caption{\emph{BBSD}}
\label{table:depparamsBBSDSS}
\end{minipage}
\end{table}

\clearpage 

\section{Conclusions and further work}
\label{sec:Conclusions}
Based on the concept of self-decomposability, in this paper we have presented a new method  to build dependent stochastic processes that are, at least, marginally \Levy . We have  developed the theoretical setting and we have shown how \sd\ subordinators can be built starting from \sd\ laws which are also infinitely divisible. Such processes are extremely useful if one wants to model such markets in which, whenever an event shocks one asset, after a certain random time delay, one can observe the effect spreading to the other ones. Applying this technique, we have embedded this feature inside some recent works based on multivariate subordinators presented by \citet{Semeraro2008}, \citet{SL2010} and \citet{BB2013} and we have shown how explicit expressions for the \chf\ and the correlation can be derived. These results are instrumental to design Monte Carlo schemes and Fourier techniques employed  to calibrate the models to real data in energy markets and to price Cross Border and Spark Spread options. We focused on German and French power and gas forward markets and we calibrated our models using a two steps calibration technique, consisting in fitting firstly marginal parameters on quoted vanilla products and secondly, the correlation on historical realizations. Numerical experiments have shown that our proposed models can catch even extreme values of correlation between assets. 
\par Our approach, and the relative developed numerical techniques, have been applied to energy markets with two correlated underlying assets only. Nevertheless, our modeling framework is very general and can be applied to an arbitrary number of underlying assets. Moreover, such a framework can be used, for example, in equity derivatives, with an arbitrary number of stocks, or in credit risk to model a chain of defaults caused by a common market shock that propagates across markets. 

On the other hand, from a more mathematical perspective some points are still open and will be the objective of future inquires. For instance, our models have \Levy\ margins but  it is still unclear whether the couple is  still a \Levy\ process. 

In addition, although most of our results are general, we focused on \sd\ Gamma subordinators. 
It will be worthwhile investigating the case of Inverse Gaussian processes, and therefore Normal Inverse Gaussian processes, in more detail, deriving for instance, an efficient Monte Carlo algorithm to simulate the relative \arem\  where some intuition may come from the results in \citet{DQZ18}. Finally, a  topic deserving
further investigation is the time-reversal simulation of such processes in order to efficiently price other contracts like swings and storages via backward simulation as detailed in \citet{PellegrinoSabino15} and \citet{Sabino20}.

\appendix
\section{Proofs}
\label{sec:AppendixA}
\subsection[Proof]{Proof of Proposition~\ref{prop:jointChfSemeraro} (See page~\pageref{prop:jointChfSemeraro})}
\begin{proof}
Substituting the expression of $Y_{j}\left(t\right)$, conditioning with respect $G_{j}\left(t\right)$ and since $W_{j}\left(t\right)$ are independent we get:
\begin{equation*}
\begin{split}
\phi_{\boldsymbol{Y}\left(t\right)}\left(\boldsymbol{u}\right)  = & \mathbb{E}\left[e^{i\langle \boldsymbol{u},\boldsymbol{Y}\left(t\right) \rangle}\right] = \mathbb{E}\left[e^{i u_{1} Y_{1}\left(t\right)+ i u_{2} Y_{2}\left(t\right) }\right] \\
 = & \mathbb{E}\left[e^{i\left(u_{1}\mu_{1} + i \frac{\sigma_{1}^{2} u_{1}^{2}}{2}\right) G_{1}\left(t\right) } e^{i\left(u_{2}\mu_{2} + i \frac{\sigma_{2}^{2} u_{2}^{2}}{2}\right) G_{2}\left(t\right) }\right]
\end{split}
\end{equation*}
Using the definition of $G_{j}\left(t\right)$ we have: 

\begin{equation*}
\begin{split}
\phi_{\boldsymbol{Y}\left(t\right)}\left(\boldsymbol{u}\right)  & =  \mathbb{E}\left[
e^{i\left(u_{1}\mu_{1} + i \frac{\sigma_{1}^{2} u_{1}^{2}}{2}\right)I_{1}\left(t\right) }
e^{i\left(u_{2}\mu_{2} + i \frac{\sigma_{2}^{2} u_{2}^{2}}{2}\right)I_{2}\left(t\right) }
e^{i\left(u_{2}\mu_{2} + i \frac{\sigma_{2}^{2} u_{2}^{2}}{2}\right)\alpha_{2}Z_{a}\left(t\right) } \right. \\
& \left. e^{i\left(\left(u_{1}\mu_{1} + i \frac{\sigma_{1}^{2} u_{1}^{2}}{2}\right) \alpha_{1} + \left(u_{2}\mu_{2} + i \frac{\sigma_{2}^{2} u_{2}^{2}}{2}\right) \alpha_{2}a\right)H_{1}\left(t\right) }
\right]
\end{split}
\end{equation*}
and, observing that $I_{j}\left(t\right)$, $H_{1}\left(t\right)$ and $Z_{a}\left(t\right)$, are mutually independent the thesis follows.
\end{proof}

\subsection[Proof]{Proof of Proposition~\ref{prop:correlationSemeraro} (See page~\pageref{prop:correlationSemeraro})}

\begin{proof}
We have to compute:

\begin{equation}
 cov\left(Y_{1}\left(t\right),Y_{2}\left(t\right)\right) = \mathbb{E}\left[Y_{1}\left(t\right)Y_{2}\left(t\right)\right] - \mathbb{E}\left[Y_{1}\left(t\right)\right]\mathbb{E}\left[Y_{2}\left(t\right)\right]
 \nonumber
\end{equation}
Substituting the expressions of $Y_{j}\left(t\right)$ and $G_{j}\left(t\right)$ and observing that 
\begin{equation*}
\mathbb{E}\left[H_{1}\left(t\right)H_{2}\left(t\right)\right]=aVar\left[H_{1}\left(t\right)\right]
\end{equation*}
the thesis follows from straightforward computations.
\end{proof}

\subsection[Proof]{Proof of Proposition~\ref{prop:jointChfSemeraroLuciano} (See page~\pageref{prop:jointChfSemeraroLuciano})}
\begin{proof}
Rewrite $\boldsymbol{Y}^{\rho}\left(t\right)$ as:

\begin{equation*}
\boldsymbol{Y}^{\rho}\left(t\right)= \boldsymbol{Y}_{\boldsymbol{I}\left(t\right)}+ \boldsymbol{Y}_{\boldsymbol{H}\left(t\right)}
\end{equation*}
where:

\begin{equation*}
\boldsymbol{Y}_{\boldsymbol{I}}\left(t\right) = 
				\left(
\begin{array}{ll}
				\mu_{1} I_{1}\left(t\right) + \sigma_{1}W_{1}\left(I_{1}\left(t\right)\right) \\
				\mu_{2} I_{2}\left(t\right) + \sigma_{2}W_{2}\left(I_{2}\left(t\right)\right)
\end{array}\right)
\end{equation*}
and:

\begin{equation}
\boldsymbol{Y}_{\boldsymbol{H}}\left(t\right) = 
				\left(
				\begin{array}{ll}
				\alpha_{1}\mu_{1}H_{1}\left(t\right) + \sqrt{\alpha_{1}}\sigma_{1}W_{1}^{\rho}\left(H_{1}\left(t\right)\right) \\
\alpha_{2}\mu_{2}H_{2}\left(t\right) + \sqrt{\alpha_{2}}\sigma_{2}\left(W_{2}^{\rho}\left(aH_{1}\left(t\right)\right) + \tilde{W}\left(Z_{a}\left(t\right)\right)\right)
 				\end{array}\right)
\nonumber
\end{equation}
The characteristic function is given by:

\begin{equation}
\begin{split}
\phi_{\boldsymbol{Y}\left(t\right)^{\rho}}\left(\boldsymbol{u}\right) = & \E\left[e^{i\langle \boldsymbol{u},\boldsymbol{Y}^{\rho}\left(t\right) \rangle}\right]  =  \E\left[e^{i\langle \boldsymbol{u},\boldsymbol{Y}_{I}\left(t\right) + \boldsymbol{Y}_{H}\left(t\right)  \rangle}\right] \\
= & \E\left[e^{i\langle \boldsymbol{u},\boldsymbol{Y}_{\boldsymbol{I}}\left(t\right) \rangle}\right]\E\left[e^{i\langle \boldsymbol{u},\boldsymbol{Y}_{\boldsymbol{H}}\left(t\right) \rangle}\right]
\end{split}
\label{eqn:chfLSGeneral}
\end{equation}
We now compute the two last term separately. Substituting the expression of $\boldsymbol{Y}_{\boldsymbol{I}}$, conditioning respect $I_{j}\left(t\right),\; j=1,2$ and remebering that $W_{1}\left(t\right)$ and $W_{2}\left(t\right)$ are idependent we have:

\begin{equation}
\begin{split}
\E\left[e^{\langle \boldsymbol{u},\boldsymbol{Y}_{\boldsymbol{I}}\left(t\right) \rangle}\right] 
= & \E\left[e^{i\left(u_{1}\mu_{1} + \frac{i}{2}u_{1}^{2}\sigma_{1}^{2}\right)I_{1}\left(t\right)}\right] \E\left[e^{i\left(u_{2}\mu_{2} + \frac{i}{2}u_{2}^{2}\sigma_{2}^{2}\right)I_{2}\left(t\right)}\right] \\
=& \phi_{I_{1}\left(t\right)}\left(u_{1}\mu_{1} + \frac{i}{2}\sigma_{1}^{2}u_{1}^{2}\right) \phi_{I_{2}\left(t\right)}\left(u_{2}\mu_{2} + \frac{i}{2}\sigma_{2}^{2}u_{2}^{2}\right)
\end{split}
\label{eqn:LSfirstCHF}
\end{equation}
Following the same approach we can compute the second term, obtaining:

\begin{equation*}
\begin{split}
\E\left[e^{\langle \boldsymbol{u},\boldsymbol{Y}_{\boldsymbol{H}}\left(t\right) \rangle}\right] = & \E\left[\E\left[e^{iu_{1}\alpha_{1}\mu_{1}H_{1}\left(t\right) +iu_{1}\sqrt{\alpha_{1}}\sigma_{1}W_{1}^{\rho}\left(H_{1}\left(t\right)\right) + 
iu_{2}\alpha_{2}\mu_{2}aH_{1}\left(t\right) +iu_{2}\sqrt{\alpha_{2}}\sigma_{2}W_{2}^{\rho}\left(aH_{1}\left(t\right)\right) }|H_{1}\left(t\right)\right]  \right . \\
& \left.\E\left[e^{iu_{2}\alpha_{2}\mu_{2}Z_{a}\left(t\right) +i u_{2} \sqrt{\alpha_{2}}\sigma_{2}\tilde{W}\left(Z_{a}\left(t\right)\right)}|Z_{a}\left(t\right)\right] \right]
\end{split}
\end{equation*}
Now we compute the inner expected values separately. The second inner expected value is:

\begin{equation*}
\E\left[e^{iu_{2}\alpha_{2}\mu_{2}Z_{a}\left(t\right) +i u_{2} \sqrt{\alpha_{2}}\sigma_{2}\tilde{W}\left(Z_{a}\left(t\right)\right)}|Z_{a}\left(t\right)\right] = e^{i\left(u_{2}\alpha_{2} + \frac{i}{2}u_{2}^{2}\alpha_{2}\sigma_{2}\right)Z_{a}\left(t\right)}
\end{equation*}
For the second therm we have that, since $H_{1}\left(t\right)$ is known:

\begin{equation*}
\begin{split}
\E\left[e^{iu_{1}\alpha_{1}\mu_{1}H_{1}\left(t\right) +iu_{1}\sqrt{\alpha_{1}}\sigma_{1}W_{1}^{\rho}\left(H_{1}\left(t\right)\right) + 
iu_{2}\alpha_{2}\mu_{2}aH_{1}\left(t\right) +iu_{2}\sqrt{\alpha_{2}}\sigma_{2}W_{2}^{\rho}\left(aH_{1}\left(t\right)\right) }|H_{1}\left(t\right)\right]  \\
 = e^{iu_{1}\alpha_{1}\mu_{1}H_{1}\left(t\right) + 
iu_{2}\alpha_{2}\mu_{2}aH_{1}\left(t\right)}\E\left[e^{iu_{1}\sqrt{\alpha_{1}}\sigma_{1}W_{1}^{\rho}\left(H_{1}\left(t\right)\right)  +iu_{2}\sqrt{\alpha_{2}}\sigma_{2}W_{2}^{\rho}\left(aH_{1}\left(t\right)\right) }|H_{1}\left(t\right)\right] 
\end{split}
\end{equation*}
The only unknown terms is the expected value. We have that:

\begin{equation*}
\E\left[e^{iu_{1}\sqrt{\alpha_{1}}\sigma_{1}W_{1}^{\rho}\left(H_{1}\left(t\right)\right)  +iu_{2}\sqrt{\alpha_{2}}\sigma_{2}W_{2}^{\rho}\left(aH_{1}\left(t\right)\right) }|H_{1}\left(t\right)\right]
=  e^{-\frac{1}{2}u_{1}^{2}\alpha_{1}\sigma_{1}^{2}\left(1-a\right)H_{1}\left(t\right)} e^{-\frac{1}{2}a\boldsymbol{u}^{T}a\Sigma\boldsymbol{u} H_{1}\left(t\right)}
\end{equation*}

where 
\[\Sigma =
\begin{bmatrix}
	\alpha_{1}\sigma_{1}^{2} & \sqrt{\alpha_{1}\alpha_{2}}\sigma_{1}\sigma_{2}\rho \\
	\sqrt{\alpha_{1}\alpha_{2}}\sigma_{1}\sigma_{2}\rho & \alpha_{2}\sigma_{2}^{2}
\end{bmatrix} 
\]
and $\boldsymbol{u} = \left[u_{1},u_{2}\right]$. Setting $\boldsymbol{\mu} = \left[\alpha_{1}\mu_{1},a\alpha_{2}\mu_{2}\right]$ we can conclude that:
\begin{equation}
\E\left[e^{\langle \boldsymbol{u},\boldsymbol{Y}_{\boldsymbol{H}}\left(t\right) \rangle}\right] =  \phi_{Z_{a}\left(t\right)}\left(u_{2}\alpha_{2} + \frac{i}{2}u_{2}^{2}\alpha_{2}\sigma_{2}\right)\phi_{H_{1}\left(t\right)}\left(\boldsymbol{u}^{T}\boldsymbol{\mu}+\frac{i}{2}u_{1}^{2}\alpha_{1}\sigma_{1}^{2}\left(1-a\right)+\frac{i}{2}a\boldsymbol{u}^{T}a\Sigma\boldsymbol{u} \right)
\label{eqn:LSsecondCHF}
\end{equation}
Using \eqref{eqn:LSfirstCHF} and \eqref{eqn:LSsecondCHF} in \eqref{eqn:chfLSGeneral} we have the thesis.
\end{proof}

\subsection[Proof]{Proof of Lemma~\ref{lem:01} (See page~\pageref{lem:01})}

\begin{proof}
Replacing the definition of $R_{1}\left(t\right)$ and $R_{2}\left(t\right)$ we get:
\begin{equation*}
\begin{split}
\phi_{\boldsymbol{R}\left(t\right)}\left(\boldsymbol{u}\right) = & \E\left[e^{iu_{1}R_{1}\left(t\right) + iu_{2}R_{2}\left(t\right)}\right] \\
= & \E\left[e^{iu_{1}\beta_{R_{1}}H_{1}\left(t\right) +iu_{2}a\beta_{R_{1}}H_{1}\left(t\right) + iu_{2}\beta_{R_{2}}Z_{a}\left(t\right) } \right.    \\          
& \left. \E\left[e^{iu_{1}\gamma_{R_{1}}W\left(H_{1}\left(t\right)\right) + iu_{2}\gamma_{R_{2}}
\left(W\left(aH_{1}\left(t\right)\right) + \tilde{W}\left(Z_{a}\left(t\right)\right)\right)}|H_{1}\left(t\right),Z_{a}\left(t\right)\right] \right]
\end{split}
\end{equation*}

\noindent We compute now the inner expected value:

\begin{equation*}
\begin{split}
& \E\left[e^{iu_{1}\gamma_{R_{1}}W\left(H_{1}\left(t\right)\right) + iu_{2}\gamma_{R_{2}}
\left(W\left(aH_{1}\left(t\right)\right) + \tilde{W}\left(Z_{a}\left(t\right)\right)\right)}|H_{1}\left(t\right),Z_{a}\left(t\right)\right] \\
& = \E\left[e^{iu_{1}\gamma_{R_{1}}W\left(H_{1}\left(t\right)\right) + iu_{2}\gamma_{R_{2}}
W\left(aH_{1}\left(t\right) \right)}|H_{1}\left(t\right)\right] \E\left[e^{iu_{2}\gamma_{R_{2}}\tilde{W}\left(Z_{a}\left(t\right)\right)}|Z_{a}\left(t\right)\right]
\end{split}
\end{equation*}

\noindent The second computation of the second expected value is immediate.

\begin{equation*}
\E\left[e^{iu_{2}\gamma_{R_{2}}\tilde{W}\left(Z_{a}\left(t\right)\right)}|Z_{a}\left(t\right)\right] = e^{-\frac{1}{2}u_{2}^{2}\gamma_{R_{2}}^{2}Z_{a}\left(t\right)}
\end{equation*}

\noindent For the first term we have:
\begin{equation*}
\E\left[e^{iu_{1}\gamma_{R_{1}}W\left(H_{1}\left(t\right)\right) + iu_{2}\gamma_{R_{2}}
W\left(aH_{1}\left(t\right) \right)}|H_{1}\left(t\right)\right]  = e^{-\frac{1}{2}\left(u_{1}^{2}\gamma_{R_{1}}^{2} +2u_{1}u_{2}\gamma_{R_{1}}\gamma_{R_{2}}a + a u_{2}^{2}\gamma_{R_{2}}^{2}\right)H_{1}\left(t\right)}
\end{equation*}

\noindent Observing that $H_{1}\left(t\right)$ and $Z_{a}\left(t\right)$ are idependent the thesis follows.
\end{proof}

\subsection[Proof]{Proof of Proposition~\ref{lem:01bis} (See page~\pageref{lem:01bis})}

\begin{proof}
Replacing the expression of $Y_{j}\; j=1,2$ we have that:
\begin{equation*}
\mathbb{E}\left[e^{\langle\boldsymbol{u},\boldsymbol{Y}\left(t\right)\rangle}\right]
=  \mathbb{E}\left[e^{iu_{1}X_{1}\left(t\right)}\right]\mathbb{E}\left[e^{iu_{2}X_{2}\left(t\right)}\right]\phi_{\boldsymbol{R}\left(t\right)}\left(\boldsymbol{a}\circ\boldsymbol{u}\right) 
\end{equation*}
Observe that, conditioning to $G_{j}\left(t\right)$, we have that:

\begin{equation*}
\E\left[e^{iu_{j}X_{j}\left(t\right)}\right] = \E\left[e^{i\left(u_{j}\beta_{j} + \frac{i}{2}u_{j}^{2}\gamma_{j}^{2}\right)G_{j}\left(t\right)}\right] = \phi_{G_{j}\left(t\right)}\left(u_{j}\beta_{j} + \frac{i}{2}u_{j}^{2}\gamma_{j}^{2}\right)
\end{equation*}
This observation jointly with Lemma \ref{lem:01} complete the proof.
\end{proof}

\subsection[Proof]{Proof of Proposition~\ref{prop:CorrelationBB} (See page~\pageref{prop:CorrelationBB})}
\begin{proof}
Computing the covariance between $Y_{1}\left(t\right)$ and $Y_{2}\left(t\right)$ we have that:
\begin{equation}
cov\left(Y_{1}\left(t\right),Y_{2}\left(t\right)\right) = a_{1}a_{2}cov\left(R_{1}\left(t\right),R_{2}\left(t\right)\right)
\label{eqn:corrProofEqn1}
\end{equation}
But, by direct computations, one can show that:
\begin{equation}
cov\left(R_{1}\left(t\right),R_{2}\left(t\right)\right) = \beta_{R_{1}}\beta_{R_{2}}a Var\left[H_{1}\left(t\right)\right] + \gamma_{R_{1}}\gamma_{R_{2}}a\mathbb{E}\left[H_{1}\left(t\right)\right]
\label{eqn:corrProofEqn2}
\end{equation}
where we used the following property:

\begin{equation*}
\E\left[W\left(H_{1}\left(t\right)\right)W\left(aH_{1}\left(t\right)\right)\right] =  a\E\left[H_{1}\left(t\right)\right]
\end{equation*}

\noindent Using \eqref{eqn:corrProofEqn1} and \eqref{eqn:corrProofEqn2} we have the thesis.
\end{proof}

\clearpage
\bibliographystyle{plainnat}
\bibliography{biblioAll}

\end{document}